\documentclass[3p,onecolumn]{elsarticle}
\usepackage{graphicx}
\usepackage{amssymb}
\usepackage{amsmath}
\usepackage{graphics}
\usepackage{color}
\usepackage[normalem]{ulem}

\journal{Communications in Nonlinear Science and Numerical Simulation}
\begin{document}

\begin{frontmatter} 

\title{Frustration induced transient chaos, fractal and riddled basins in coupled limit cycle oscillators}
\author[sas]{K.~Sathiyadevi}
\author[cnld]{S.~Karthiga}
\author[sas]{V.~K.~Chandrasekar \corref{cor1}}
\ead{chandru25nld@gmail.com}
\author[iis]{D.~V.~Senthilkumar \corref{cor2}}
\ead{skumar@iisertvm.ac.in}
\author[cnld]{M.~Lakshmanan}

\cortext[cor1]{Corresponding author} 
\cortext[cor2]{Corresponding author} 

\address [sas]{Centre for Nonlinear Science \& Engineering, School of Electrical \& Electronics Engineering, \\ SASTRA Deemed University, Thanjavur- 613 401, India.} 
\address[cnld]{Centre for Nonlinear Dynamics, School of Physics, Bharathidasan University, Tiruchirapalli-620024, India}
\address[iis]{School of Physics, Indian Institute of Science Education and Research, Thiruvananthapuram -695016, India}

\date{\today}




\begin{abstract}
This report unravels frustration as a source of transient chaotic dynamics even in a simple array of coupled limit cycle oscillators.  The transient chaotic dynamics along with the multistable nature of frustrated systems facilitates the existence of complex basin structures such as fractal and riddled basins.  In particular, we report the emergence of transient chaotic dynamics around the chaotic itinerancy region, where the basin of attraction near the chaotic region is riddled and it becomes fractal basin when we move away from this region.  
 Normally the complex basin structures are observed only for oscillatory states but  surprisingly it is also observed even for oscillation death states.  Interestingly super persistent transient chaos is also shown to emerge in the coupled limit cycle oscillators, which manifests as a stable chaos for larger networks.
\end{abstract}


\end{frontmatter}

\section{Introduction}
Frustrated systems have been receiving a great deal of interest over several years as they have become a source of multiple ground states and/or multistability  because of their inability to simultaneously minimize the competing
interaction energies between the interacting constituents \cite{spin}.   As a classical example, one can consider the case of three Ising spins with anti-ferromagnetic interaction placed at the corner of an equilateral triangle, where the inability to align each spin anti-parallel with each other results in frustration and gives rise to multiple ground states \cite{spin,bhumi,phystod}.  Further, the frustrated systems facilitate the onset of metastable states when they tend to relax themselves from frustration \cite{fru_ch1, brain}.  Such frustration induced multistable and metastable states are the pivotal factors for the underlying rich dynamics of  biological networks such as neural networks, genetic networks, slime models \cite{bhumi, fru_ch1,brain, multi, slime, slime2} and so on.  For instance, one of the metastable states, namely,  frustrated chaos (or chaotic itinerancy) observed in such systems can explain stimulus independent switching between the various cognitive states observed in the brain and is central to the information processing ability of the latter \cite{fru_ch1, brain,iti,tsu}. The multistable nature results in multiple operating regimes in  biological systems facilitating functional flexibility in response to external stimuli \cite{multi}.
\par Despite the emergence of a rich variety of dynamical behaviors due to multistable and metastable dynamics observed in frustrated systems,  frustration has not yet been reported as a source of finite time chaotic behavior, namely the transient chaotic states.  Such a transient chaotic behavior characterized by the combined features of chaotic and convergent dynamics produces complex basin structures, namely fractal and riddled basins due to the interplay with multistable nature of the system \cite{tel, awadhesh1,prx_dis,prl_2017, multi_kap_awa}.  
It was shown that the rapid transitions among different ordered states observed during cognitive tasks or memory recall are mediated by chaotic transients \cite{cerebral,moss,weakly,ssinha}.   Due to these reasons, transient chaotic associative neural networks are found to be useful in the field of neurocomputing and others.   For instance, these networks are useful for solving combinatorial optimization problems \cite{aihara, aihara2}, progressive memory recalling, temporal pattern associations \cite{lee}, maintenance-scheduling problem of generators in a practical power system and assigning channels in a cellular mobile network \cite{phone, phone2}. In view of these applications, including combinatorial optimization problem, it has been shown that the transient chaotic behavior prevents the system to be trapped from a local minimum and provide a quick search over the fractal space \cite{aihara, aihara2}.  The information about the input conditions should fade after converging to a particular ordered state in order to process the next input stimulus and hence the converging dynamics of transient chaos facilitates fading memory. Furthermore, the searching space is reduced due to fractal basin structure which favors an efficient searching for the required ordered or memory state \cite{aihara, aihara2}.  Due to these reasons, this type of transient chaotic networks combining chaotic and convergent dynamics have high ability to realize an efficient search for a variety of optimization problems concerned with neuro-computing.  The efficiency of such networks has been studied from both theoretical and experimental levels.  Experimental implementation of the scheme can be seen in \cite{ameoba} where neuro-computing system is based on amoeboid unicellular organism,  namely the true slime mold Physarum.  This system is found to show chaotic itinerant or wandering dynamics where spontaneous transition among the multiple quasi-stable modes even without any external perturbation can be seen.  It allows one to reach a desirable orbit or attractor quickly by suitable external forcing.  The results show the high capability of the system to search the optimal solution of the four-city traveling salesman problem with a high probability.   Particularly, information processing based on edge of chaos and nonlinear transient computing are  promising contemporary research topics  both from computational neuroscience and dynamical systems points of view \cite{weakly,ch_trans,app,skumar,edgge,mass}, where the edge of chaos that lies between ordered and chaotic regimes is proved to be the optimal region for computations in the cutting edge computing paradigm called liquid state machines (reservoir computing) \cite{langton}. 
\par The transient chaotic behavior has been mainly achieved in the literature by controlling chaos \cite{schol1, schol2} or by chaotic anealing \cite{ssinha,aihara,aihara2,ogy,ogy2}, while the associated complex basin structures have been largely reported in driven dissipative systems, chaotic systems, delay coupled systems and maps \cite{ fractal, trans_saba, sanjuan_trans, delay_saha, trans_dissi}.  {From the experimental realization and application point of view, simple non-chaotic systems exhibiting transient chaos along with the associated complex basin structures are preferable as any experimental error and associated noise will not get amplified across the network, due to non-chaoticity, which may result in more accurate and robust performance.}  Hence, in this report we explore the possibility of generating transient chaos and complex basin structures (including fractal and riddled basins) in a simple system of coupled limit cycle oscillators by exploiting frustration dynamics.  In addition, this report will also elucidate the possibility to inherit the chaotic wandering dynamics in a simple system of frustrated limit cycle oscillators, where such a chaotic wandering or itinerant dynamics is found to facilitate the spontaneous recollection of associated memory states in accomplishing a task.  
\par In particular, in this report, we demonstrate frustration induced transient chaos and multistability in a paradigmatic model of coupled van der Pol oscillators exhibiting limit cycle oscillations with both attractive and repulsive couplings. Importantly, the competing attractive and repulsive couplings plays a crucial role in  many biological  processes. For instance, coexistence of excitatory and inhibitory synaptic couplings can be found in a pair of neurons. In the gene regulatory network the counteracting positive and negative feedback loops are used to operate various functions including bistable switches, oscillators and excitable devices \cite{ar_app1,ar_app2,ar_app3}.  
  Hence, we also aim towards  understanding dynamical behavior due to the competing effect of attractive and repulsive couplings. In the non-frustrated case of two coupled systems ($N=2$), the system exhibits interesting multistabilities and the frustration comes into play when $N$ is increased to $3$ which results in rich dynamics in the multistable region of anti-phase oscillations with oscillation death observed for $N=2$.  Interestingly, the frustration that exists among inhomogeneous oscillations give rise to chaotic itinerancy and transient chaotic behaviors.  The transient chaotic behavior in combination with the multistability that arises due to the inhomogeneous oscillatory and oscillation death states leads to riddled and fractal basin structures.  In addition, we also observe super persisting transient chaos {(i.e. the chaotic behavior persists for very long time)} in the region where the attractive and repulsive couplings are almost of equal strengths.  This type of super persisting transient chaos has been first reported theoretically by C. Grebogi {\it et al.} in \cite{greb_new}.   With the coupled chaotic electronic circuits, the first experimental evidence of long transient chaos is reported in \cite{exp_new} and their existence has also been reported in a practical example of Tilt-A-Whirl \cite{boook_new}. The super persistent transient chaotic state is stabilized to stable chaos with the increase of network size ($N>9$) giving rise to interesting multistability between completely ordered state and disordered state.  Further, we report the emergence of fractal basins in the oscillation death region for the first time in the literature.  We have also extended our analysis to a network of $N=100$ oscillators in search of extreme multistability with rich dynamical behavior.  

The rest of the paper is organized as follows: In Sec.~\ref{S:2}, we introduce our model of globally coupled van der Pol oscillators and the corresponding dynamical behavior will be discussed from smaller network to larger networks.  The observed results are summarized in Sec.~\ref{S:3}. The additional information about transient chaotic nature, Hopf bifurcation curves, mean exit time of transient chaos induced crisis, dynamics in riddled and fractal basin regions, uncertainty exponents and  the transition from transient chaotic nature to stable chaos while increasing size of the networks  will be discussed in the Appendix  A.

\section{Dynamical behavior in coupled vdP oscillators}
\label{S:2}
\subsection{Model}
\par We consider a simple paradigmatic model of the van der Pol oscillator  (vdP) which is known to be a primary model for describing self-excited oscillations observed extensively in biological, physical and engineering systems.  The FitzHugh-Nagumo model is a modified form of the vdP oscillator \cite{vdp_chaos}. From a practical point of view,  van der Pol oscillators can be implemented using appropriate electronic circuit \cite{vdp_elec}.  The set of coupled van der Pol oscillators is represented as
\begin{eqnarray}
\dot{ x}_{i} &=& { y}_{i} + \epsilon_1({\overline{x}}-{x}_{i}), \qquad i=1,2,3,...N, \nonumber \\ 
\dot{ y}_{i} &=&\alpha (1-{ x}^2_{i}) { y}_{i} -{ x}_{i} -\epsilon_2({ \overline{y}}-{ y}_{i}).
\label{model}
\end{eqnarray}
where, $x_i$, $y_i$ are the state variables of the $i^{\mathrm{th}}$ system and $\alpha$ is the strength of the linear (negative) and the nonlinear (positive) damping terms, while $\epsilon_1$ and $\epsilon_2$ denote the strengths of the attractive and the repulsive couplings, respectively.   Throughout this report, $\alpha$ is taken to be $0.5$.  In Eq. (\ref{model}), ${ \overline{x}} = \displaystyle{\frac{1}{N} \sum \limits_{i=1}^{N} { x}_{i}}$, ${ \overline{y}} = \displaystyle{\frac{1}{N}\sum \limits_{i=1}^{N}{ y}_{i}}$.  The dynamical behavior are obtained by solving the system of equation (\ref{model}) using  Runge Kutta $4^{\mathrm{th}}$ order integration method with step size $h=0.01$. We also wish to note that there is no need to restrict $h=0.01$, and one may consider other suitable step sizes, for example $h = 0. 005$ or $0.05$,  for the numerical integration and the results are reproducible for those values of $h$ too. 
\subsection{Non-frustrated case (N=2)}

\par A minimal network that can facilitate the frustration dynamics is a network of three coupled oscillators.  Prior to unraveling the dynamical transitions in a three coupled network of vdP oscillators, we will briefly summarize the dynamics of two coupled vdP oscillators.  When $N=2$, the attractive interaction tends to align the coupled oscillators to exhibit in-phase oscillations, while the repulsive coupling has the tendency to align them to evolve in anti-phase with each other.  The in-phase synchronized state (IPS) is stable for appreciably large attractive coupling ($\epsilon_1$) whereas the anti-phase synchronized (APS) state is predominant for a large repulsive coupling ($\epsilon_2$) as is evident from  Fig. \ref{three}(a).  Multistability between APS and IPS states is also observed in the region $R_1$ (see Fig. \ref{three}(a)) for smaller values of  $\epsilon_1$ and $\epsilon_2$ as a result of the  trade-off between the attractive and repulsive couplings.  Oscillation death is also found to occur in the two coupled system for higher values of $\epsilon_1$ and $\epsilon_2$ where we have observed the stabilization of the system towards one of the inhomogeneous steady states, 
\begin{eqnarray}
x_1^*=-x_2^*= \pm \sqrt{\frac{\alpha \epsilon_1-1+\epsilon_1 \epsilon_2}{\alpha \epsilon_1}}, \quad
y_1^*=-y_2^*= \epsilon_1 x_1^*. \label{22od}
\end{eqnarray}
Linear stability analysis elucidates that this anti-symmetric state is stable in the region $\epsilon_1>\frac{1}{\epsilon_2}$ when $\epsilon_2<1.0$ and in the region $\epsilon_1>1.0$ when $\epsilon_2 \geq 1.0$ while the stabilization of the inhomogeneous (OD) states occur through a sub-critical Hopf bifurcation. Also the  OD state coexists with the APS and IPS  states to give rise to bistability in the regions $R_2$ and $R_3$, respectively, while multistability among them arises in $R_4$.  The analytical  stability curve for OD state  is shown in Fig.~\ref{three}(a). Also, the boundary of  in-phase synchronized state is  found through master stability analysis (see Fig. \ref{three}(a)). 
%
\subsection{Frustrated case (N=3)}
\par
\begin{figure}[!ht]
	\centering
	\includegraphics[width=1.0\linewidth]{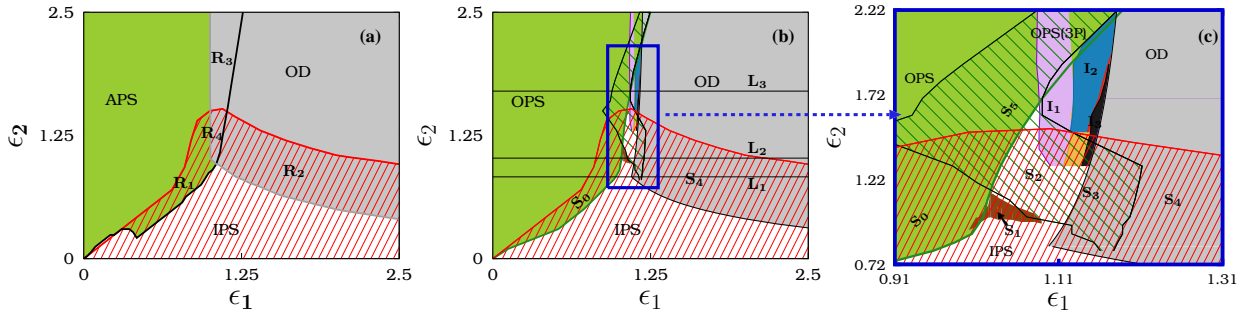}
	\caption{(a): Regions of different collective  dynamical states in the two coupled vdP oscillators. The regions $R_1$, $R_2$, $R_3$ and $R_4$, respectively, show the multistability between the states IPS-APS, IPS-OD, APS-OD and IPS-APS-OD. Fig. (b)Regions of different collective states from the three coupled vdP oscillators and Fig. (c) is the enlarged portion of a part of Fig. (b).  In Figs. (b) and (c), the multistable regions are represented by $S_0$ (IPS-OPS multistable region), $S_2$ (OPS(3P)-IPS), $S_3$ (OPS(3P)-IPS-OD), $S_4$ (IPS-OD), $S_5$ (OPS(3P)-OPS).  Super persisting transient chaos has been observed in the region $S_1$.  $I_1$ and $I_3$ correspond to the stable region of H-IN and IN states, respectively.  Chaotic switching behavior is observed in the region $I_2$. The boundaries of IPS and OD are the anlytical stability curves obtained through master stability function and	linear stability analysis.} 
	\label{three}	
\end{figure}

 On adding one more element to the system of two coupled vdP oscillators, We find the attractive coupling  has the tendency to force all the three oscillators to exhibit in-phase oscillations whereas the phase-repulsive coupling facilitates the three oscillators to align in anti-phase with  each other in order to minimize the energy.  But this sort of simultaneous alignment among the three oscillators  showing anti-phase  nature to each other is practically impossible and as a consequence geometric frustration comes into play.   Thus in the presence of the repulsive coupling, instead of anti-phase alignment (with $\pi$ phase difference), the three oscillators are aligned in an out-of-phase synchronized state (OPS) with $\frac{2 \pi}{3}$ phase difference  which is also called as the rotating wave \cite{ro_wave}.   The different phase synchronization including tri-phase synchronization  are also found in a frustrated system of calling frogs \cite{cf}.   The stable regions of the IPS and OPS states in a system of three coupled vdP oscillators are depicted in Fig. \ref{three}(b).  Further, the OD state observed in the three coupled oscillators is no more anti-symmetric ($x_i \neq -x_j$, $i,j \in \{1,2,3\}$ and $i \neq j$) and it takes either of the forms given by $B$ and $C$ below.\\
OD State - $B$:
%
\begin{eqnarray}
&\mathrm{B_1}:& x_k^*=+ \frac{(a(1-{x_i^*}^2)+b-3)x_i^*}{a(1-{x_i^*}^2)+b}, \quad y_k^*=-2 y_i^* \label{od2} \nonumber \\
&\mathrm{B_2}:& x_i^*=x_j^*=+ \sqrt{\frac{a+b-z}{a}}, \quad y_i^*=y_j^*=\epsilon_1(\frac{x_i^*-x_k^*}{3}), \quad i,j,k \in \{1,2,3\}, \quad i = j \neq k.   \label{od1} 
\end{eqnarray}
%
OD State - $C$:
\begin{eqnarray}
&\mathrm{C_1}:&x_k^*= \frac{(a(1-{x_i^*}^2)+b-3)x_i^*}{a(1-{x_i^*}^2)+b}, \quad y_k^*=-2 y_i^* \label{od4} \nonumber \\ 
&\mathrm{C_2}:&x_i^*=x_j^*=- \sqrt{\frac{a+b-z}{a}}, \quad y_i^*=y_j^*=\epsilon_1(\frac{x_i^*-x_k^*}{3}), \quad i,j,k \in \{1,2,3\}, \quad i = j \neq k.  \label{od3} 
\end{eqnarray}

where, $z=\frac{5}{3}-\frac{2^\frac{1}{3}p}{3(q+\sqrt{q^2+4p^3})^{\frac{1}{3}}}+\frac{(q+\sqrt{q^2+4p^3})^{\frac{1}{3}}}{3 \times 2^{\frac{1}{3}}} $ with $p=-7+12a+12b$, $q=-20-18a-18b$, $a=\epsilon_1 \alpha$ and $b=\epsilon_1 \epsilon_2$.  Depending on the initial conditions, the system takes one of the steady state configurations given by $B$ or $C$.   The subscripts of $B$ and $C$ in Eqs. (\ref{od1})-(\ref{od3}) denote the number of oscillators that can populate the corresponding steady state configuration.  In particular, if $B$ is the asymptotic state of the system then as $t \rightarrow \infty$, one among the three oscillators reach the state $B_1$ while the other two oscillators populate the state $B_2$.  Considering the commutation symmetry ($x_i \leftrightarrow x_j$), of the system, totally six possible stable steady state configurations can be observed in the gray region (marked as OD) in Fig. \ref{three}(b). 
\par Comparing the stable regions of IPS, OPS and OD states in the case of three coupled oscillators with that of the two coupled oscillators case in Fig. \ref{three}(a), significant changes and novel dynamics can be observed in the multistable regions of OD and APS states (that is, corresponding to regions $R_3$ and $R_4$ in Fig. \ref{three}(a)).  Figure \ref{three}(b) shows that in the latter parametric range, the multistability between OPS and OD states is largely suppressed with the emergence of rich dynamical states including three periodic OPS (OPS(3P)), inhomogeneous oscillatory, and chaotic states.   The stable regions of these states are also shown clearly in Fig. \ref{three}(c).  The dynamical regimes between the two coupled and three coupled oscillators are almost identical for weaker attractive and repulsive  couplings.  In such parametric ranges, we observe transition from OPS state to IPS state through a multistable region of OPS and IPS (region $S_0$) as a function of $\epsilon_1$.  Increasing the strength of $\epsilon_1$ further leads to the stabilization of OD state and a bifurcation analysis elucidates that it is stabilized via a subcritical Hopf bifurcation (as in the case of $N=2$) for all values of $\epsilon_2$ that lie below the line $L_1$ in Fig. \ref{three}(b).
\subsection{Super persistent transient chaos}
\par
\begin{figure}[htb!]
	\centering
	\includegraphics[width=0.9\linewidth]{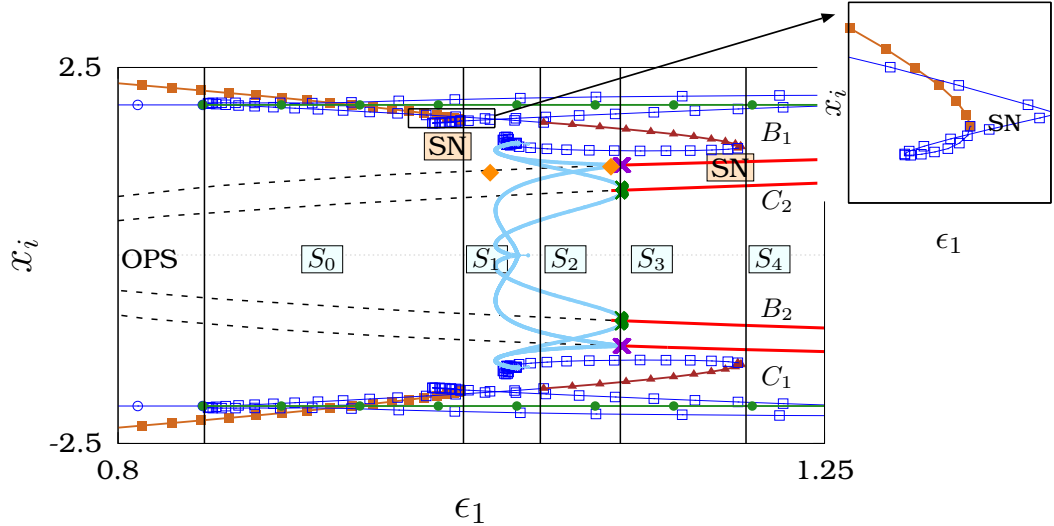}
	\caption{Bifurcation diagram for $\epsilon_2=1.0$.  In the figure, filled circle (green colored), filled square (brown curve), filled triangle (dark brown curve), respectively, show the stable nature of IPS, OPS and OPS(3P) states. Blue curve and the lines with unfilled squares and circles denote the unstable nature of the periodic orbits.  Supercritical Hopf bifurcation occurs at the boundary between $S_2$ and $S_3$, the purple and dark green colored cross (X) marks denote the bifurcation of stable inhomogeneous limit cycle oscillations.  The stable region of these inhomogeneous oscillations is narrow and they are represented by cross marks.   The orange color diamonds denote the points HB$_1$ and HB$_2$ given in Fig. \ref{zero}(c) of the  Appendix A.2.  The continuous line represents the stable nature of the OD state (it is stabilized via a supercritical Hopf bifurcation here) and dashed line indicates the unstable steady states.} 
	\label{bifur1}	
\end{figure}

\begin{figure}[htb!]
	\centering
	\includegraphics[width=0.95\linewidth]{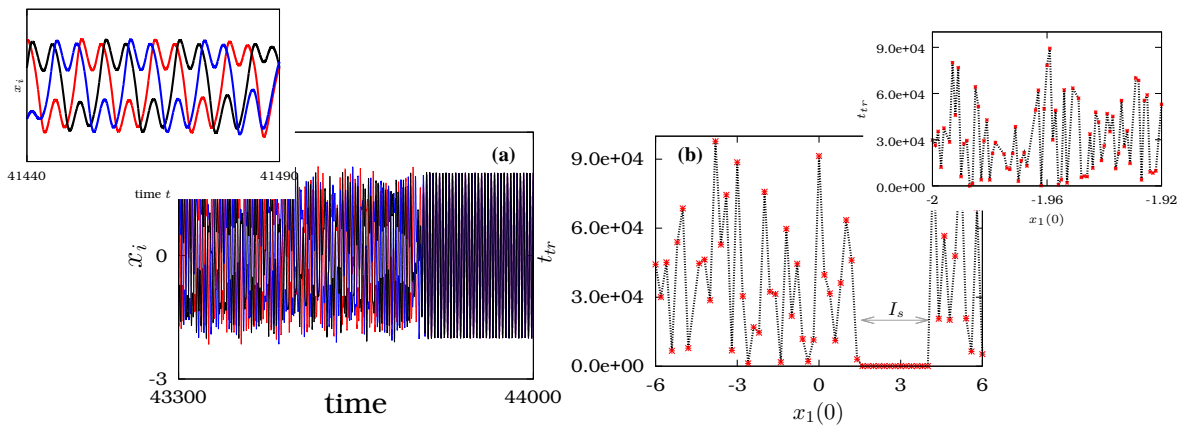}
	\caption{ Figure (a) shows the transient chaotic behavior in the region $S_1$ and the subset in Fig. (a) shows three periodic OPS(3P) state in the transient evolution. Fig. (b) and its inset display the transient times taken by the system in this region for various initial conditions corresponding to $x_1(0)$.} 
	\label{bif1}	
\end{figure}

  Surprisingly stronger repulsive coupling gives rise to the onset of chaotic transients and other non-trivial dynamics in specific parametric ranges (in the regions where APS-OD coexists in the case of two coupled vdP oscillators) in between the lines $L_2$ and $L_3$ in Fig. \ref{three}(b) for $\epsilon_2=1.0$ and $1.7$, respectively. First we will discuss  the dynamical transitions of the system as a function of $\epsilon_1$  for $\epsilon_2=1.0$,  which are illustrated in the bifurcation diagram in Fig. \ref{bifur1} (using XPPAUT software).   It elucidates  that for smaller values of $\epsilon_1$ ($\epsilon_1< 0.85$), OPS state is the only stable state (stable nature of this state is represented by lines connected by  filled (brown) squares).  By increasing the value of $\epsilon_1$, the IPS state (whose stable nature is represented by lines connected by filled (green) circles) also becomes stable at the left boundary of region $S_0$.  Thus in the region $S_0$, both OPS and IPS states are stable. {On further increasing $\epsilon_1$  the OPS state gets destabilized ( which is indicated by lines connecting unfilled squares) through a fold (or saddle-node like) limit cycle bifurcation (the bifurcation point is represented as SN in Fig. \ref{bifur1}) where the stable limit cycle collides with the unstable ones and they get annihilated via fold limit cycle.  The occurrence of fold limit cycle bifurcation is further illustrated via the subset of Fig. \ref{bifur1}.}   After the destabilization of the OPS state, the same oscillatory (unstable) branch, undergoes folding and stabilizes as a three periodic OPS (OPS(3P)) state (stable nature of it is denoted by filled brown colored triangles) in the regions $S_2$ and $S_3$.  In this OPS(3P) state, the oscillators are aligned with a $\frac{2 \pi}{3}$ phase difference (and its temporal behavior looks like the one given in the subset of Fig. \ref{bif1}).
\par  In the above case, the OPS(3P) state is not stabilized soon after the destabilization of OPS state and so in the intermediate region (namely, $S_1$), IPS state is the only stable state.  It is to be noted that these initial conditions lead to stable OPS states in the region $S_0$ which will also lead to stable IPS state asymptotically in the parametric range of $S_1$, while the transient dynamics in this region non-trivially shows longer time transient chaotic behavior or super persistent transient chaos as shown in Fig. \ref{bif1}(a).  {The unstable orbits present in the region $S_1$ constitute a chaotic saddle which results in the long transient chaotic evolution.   The fold or saddle-node type limit cycle bifurcations that occur in the $S_0$ and $S_3$ regions may be responsible for the emergence of such chaotic saddle.}   One can also observe the short intervals of OPS(3P) states among the super persistent chaotic dynamics as illustrated in the inset of Fig. \ref{bif1}(a) and it mimics intermittent like behaviors in transient chaotic evolutions.  Such an intermittent transient chaotic nature is more pronounced if one examines the dynamics near the boundary of the region $S_2$. {The chaotic nature of the transient dynamics is confirmed using $0-1$ test and the details of which are illustrated in Appendix A.1.  The transient chaotic nature can also be corroborated from the transient time taken by the system corresponding to different initial conditions of depicted in Fig. \ref{bif1}(b).}  In this figure, we have fixed $y_1(0)=0.4$, $x_2(0)=1.8$, $y_2(0)=0.2$, $x_3(0)=2.0$, $y_3(0)=0.1$ and calculated the transient times $t_{tr}$ for different values of $x_1(0)$, where the non-smooth variation of $t_{tr}$ with respect to various initial conditions  elucidate the presence of chaotic saddles.  Further, it is also evident from the sensitive dependence of $t_{tr}$ with respect to the initial conditions as illustrated in the subset of Fig. \ref{bif1}(b).  This transient chaotic state is not only observed in the region between the stable arena of OPS and OPS(3P) states (region $S_1$)  in Fig. \ref{three}(b) but also near the boundaries of stable regions of the OPS state (near the boundary of $S_0$) and OPS(3P) state (near the boundary of $S_2$). But in the latter regions, the transient chaotic dynamics exists only for a finite time and super persisting chaotic transients are observed only in the region $S_1$. 

\par After the parametric regions with transient chaos and stable OPS(3P) states, Fig. \ref{bifur1} displays the existence of stable OD state for larger values of $\epsilon_1$ (regions $S_3$ and $S_4$).  In Fig. \ref{bifur1}, we observe four branches of lines ($B_1$, $B_2$, $C_1$ and $C_2$) which correspond to the two steady state configurations of $B$ and $C$  in Eqs. (\ref{od1})-(\ref{od3}).  Depending on the initial condition, either of the steady state configurations $B$ or $C$  will be  stabilized.  If $B$ ($C$) is the asymptotic state, two oscillators will settle at the value given by $B_2$ ($C_2$) while the other takes up the value $B_1$ ($C_1$).  Figure \ref{bifur1} shows that the stabilization of these inhomogeneous steady states occur via the supercritical Hopf bifurcation in contrast to the one observed for lower values of $\epsilon_2$ (and also in the case of two coupled system) where OD occurs via a sub-critical Hopf bifurcation.  The parametric ranges over which the OD state is stabilized through sub-critical and super-critical Hopf bifurcations are shown in the Appendix A.2. (see Fig. \ref{zero}(c)).  The inhomogeneous limit cycles resulting from this supercritical Hopf bifurcation are found in a very narrow range of the parametric space as shown in Fig. \ref{bifur1} {(also in the Figs. \ref{three}(b) and \ref{three}(c))}.  


\subsection{Chaotic itinerancy and fractal basins}
\begin{figure}[h]
	\centering
	\includegraphics[width=0.8\linewidth]{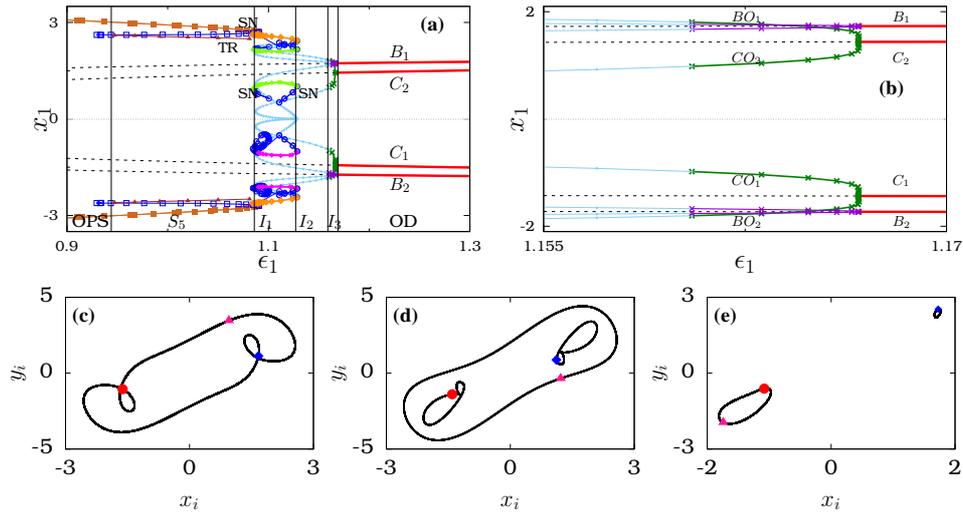}
	\caption{Fig. (a) shows the bifurcation diagram for $\epsilon_2=1.7$.  In the figure, filled circle (green colored), filled square (brown curve), filled triangle (dark brown curve) respectively show the stable nature of IPS, OPS and OPS(3P) states.  Blue dotted curves and the ones with unfilled squares and circles denote the unstable nature of the periodic orbits.  The green and pink curves with filled diamonds represents the stable inhomogeneous oscillatory branches and the orange curve with diamonds represents the stable homogeneous oscillatory branch observed in the H-IN (or $I_1$) region.  The purple and green curves with cross (X) marks in region $I_3$ represent the stable nature of inhomogeneous oscillations that emerge out of the supercritical Hopf bifurcation.  The continuous red line represents the stable nature of the OD state and dashed line shows the unstable steady states. Fig. (b) Zoomed view of the supercritical Hopf bifurcation seen in Fig. (a). Figs. (c), (d) and (e) show the phase space trajectories of OPS(3P), H-IN and IN states.  } 
	\label{bif2}	
\end{figure}
\par  In addition to the super persistent transient chaos, other interesting dynamics including inhomogeneous oscillations and stable and finite time transient chaos can be observed in the region between OPS and OD states.   To illustrate these, we have fixed the strength of the repulsive coupling as $\epsilon_2=1.7$ and the corresponding bifurcation diagram along the line $L_3$ in Fig. \ref{three}(b) is depicted in Figs. \ref{bif2}(a) and \ref{bif2}(b) ( which is a closer view of Fig. \ref{bif2}(a)).   As discussed for $\epsilon_2=1.0$, Fig. \ref{bif2}(a) depicts the fact   that  the OPS state, which is stable for lower values of $\epsilon_1$, becomes unstable through a fold (saddle-node like) limit cycle bifurcation as a function of $\epsilon_1$. The OPS(3P) state is stabilized even before the destabilization of OPS state and so there is no void between the out-of phase synchronized states (that is, OPS and OPS(3P)) as a function of  $\epsilon_1$ in contrast to the one observed in Fig. \ref{bifur1}.  The phase space trajectory of OPS(3P) state is shown in Fig. \ref{bif2}(c).   Increasing $\epsilon_1$, two inhomogeneous oscillatory branches (shown by green and pink curves with filled diamonds) and a homogeneous oscillatory branch (shown by orange colored lines with unfilled diamonds) are found to be stabilized in the region $I_1$.  The dynamics of the system in this region of homogeneous and inhomogeneous (H-IN) oscillations is shown in Fig. \ref{bif2}(d) which elucidates that the three oscillators are distributed in different orbits.  Even though the three oscillators are in different periodic orbits, they oscillate with the  same frequency and are phase synchronized.  Taking into account the commutation symmetry of the system, six different configurations are possible in this particular state.

\par After the destabilization of H-IN state, there are no stable states in the region $I_2$ as evident from Fig. \ref{bif2}(a).   Before getting into the details of dynamics in the  region $I_2$, we first discuss the dynamics observed in the region $I_3$ where the inhomogeneous (IN) oscillations emerge from the supercritical Hopf bifurcation.  For a better visualization of this bifurcation and the oscillatory branches ($BO_1$, $BO_2$, $CO_1$ and $CO_2$ ) that emerge from the steady state branches, a portion of Fig. \ref{bif2}(a) is enlarged in Fig. \ref{bif2}(b). A closer look at the dynamics in this range of  $\epsilon_1$ reveals that the inhomogeneous periodic orbits follow the same scenario followed by the inhomogeneous steady states from which they are bifurcated.  In other words, similar to  the OD states, the inhomogeneous periodic orbits can exist in two possible configurations, namely, $BO$ or $CO$.  If $BO$ ($CO$) is asymptotically stable then any two of the oscillators converge to the orbit $BO_2$ ($CO_2$) (see Fig. \ref{bif2}(b)) while the third oscillator will be designated to the orbit $BO_1$ ($CO_1$) as shown in Fig. \ref{bif2}(b).   
Six different configurations are possible in this IN state giving rise to multistability in the region $I_3$.   Figure \ref{bif2}(e) illustrates the dynamics  of the  three coupled vdP oscillators in the region $I_3$ for a particular initial condition.  It shows that two of the oscillators evolve in the same periodic orbit (namely, $BO_2$) with a $\pi$ phase (anti-phase) difference while the third oscillator occupies a separate orbit (namely, $BO_1$).  The frequencies of all the three oscillators are found to be the same  in this IN state despite their amplitudes and shapes of the periodic orbits are not symmetric.  It is interesting to note that in order to accommodate two oscillators, the orbits $BO_2$ and $CO_2$ are having larger amplitudes while that of $BO_1$ and $CO_1$ are smaller.  This strongly corroborates an interesting form of spontaneous symmetry breaking where the identical set of oscillators having commutation symmetry exhibit heterogeneous dynamics. It is to be noted that this symmetry breaking is also evident in homogeneous and inhomogeneous oscillating states (see Fig. \ref{bif2}(d)).

\begin{figure}[htb!]
	\centering
	\includegraphics[width=0.85\linewidth]{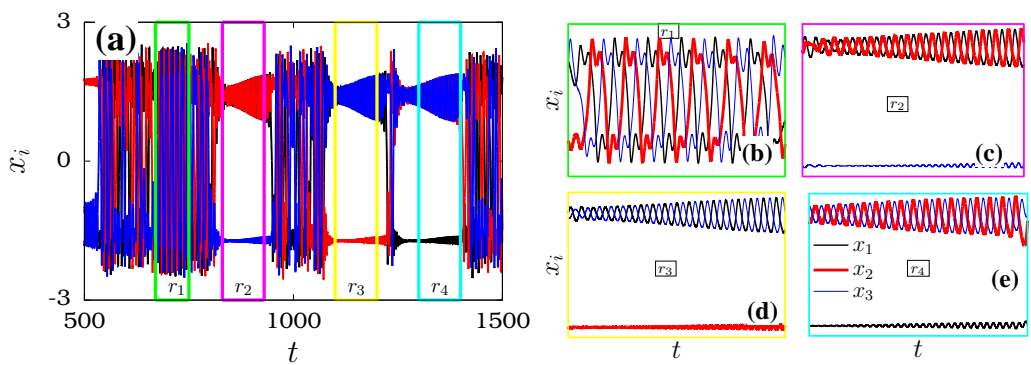}
	\caption{(a) chaotic itinerancy observed in region $I_2$ elucidating the switching between different attractors of the system.  Dynamics in the intervals $r_1$, $r_2$, $r_3$ and $r_4$ are displayed in Figs. (b), (c), (d) and (e), respectively. The time trajectory of the system in the interval $r_1$ is similar to OPS(3P) oscillations, while in the intervals $r_2$, $r_3$ and $r_4$ it  looks similar to the ones observed in the  region $I_3$ given in Fig. \ref{bif2}.}
	\label{swi}
\end{figure}
\par The frustration is also prominent in the regions of OPS, OPS(3P), H-IN, IN and OD states where the system cannot accommodate all the three oscillators in anti-phase with each other.  In the IN state, two of the oscillators evolve in anti-phase with each other while the third oscillator cannot do so and so the frustration effects are more prominent here.  The frustrated system normally has the tendency to relax itself into a non-frustrated state and as a consequence the inhomogeneous oscillations lose their stability in the region $I_2$ and show switching between different possible states in an erratic way\cite{fru_ch1}.   For instance, we have depicted the above mentioned metastable dynamics in Fig. \ref{swi}(a) and the peculiar dynamical states observed in different time windows are  displayed in Figs. \ref{swi}(b)-\ref{swi}(e).  The oscillations around the unstable OPS(3P) state can be seen in the time interval $r_1$ (see Fig. \ref{swi}(b)) and in the time window $r_2$ two of the oscillators are in anti-phase state while the third oscillator is moving independently as illustrated in Fig. \ref{swi}(c).   In the intervals $r_3$ and $r_4$,  the first and third oscillators, and the second and third oscillators, respectively, are in anti-phase with the other two as shown in Figs. \ref{swi}(d) and \ref{swi}(e), respectively.  Such a type of erratic wandering among different unstable periodic orbits is known as chaotic itinerancy or frustrated chaos and this type of metastable dynamics is of great interest in neurodynamics where the spontaneous switching between different states enables sequential memory recalling.  The region with chaotic itinerant behavior is denoted as $I_2$ in Figs. \ref{three}(b) and \ref{three}(c) in a certain range of $\epsilon_2$ which is suppressed largely for large $\epsilon_2$.  This is because the stabilization of  the inhomogeneous steady states occur via a sub-critical Hopf bifurcation (shown in Fig. \ref{zero}(c) of Appendix A.2.) for large $\epsilon_2$ by destabilizing the stable inhomogeneous limit cycles in the corresponding ranges of $\epsilon_2$.  Due to the sub-critical Hopf bifurcation, the emerging unstable periodic orbits coexist with the stable inhomogeneous steady states and so there will not be any chaotic states for large $\epsilon_2$. 
\par Even though the above mentioned stable chaotic behavior in coupled limit cycle oscillators exist for a shorter range of parameters, its impact in the form of transient chaos is wide over the parametric ranges.  One can find the existence of transient chaos widely around the stable chaotic region (region - $I_2$), that is in the regions $I_1$, $I_3$ and near the boundaries of $S_5$ and OD regions. { Even in the present case, we observe the sudden destabilization of chaotic orbit and it does not follow any of the routes other than the existence of transient chaos, which is found to be short lived in time.  This type of transient chaos around the chaotic regions may arise due to  crisis \cite{uncert}.  In Appendix A.3., we have deduced the mean exit time scales of the transient chaos.   }
\begin{figure}[htb!]
	\centering
	\includegraphics[width=1.0\linewidth]{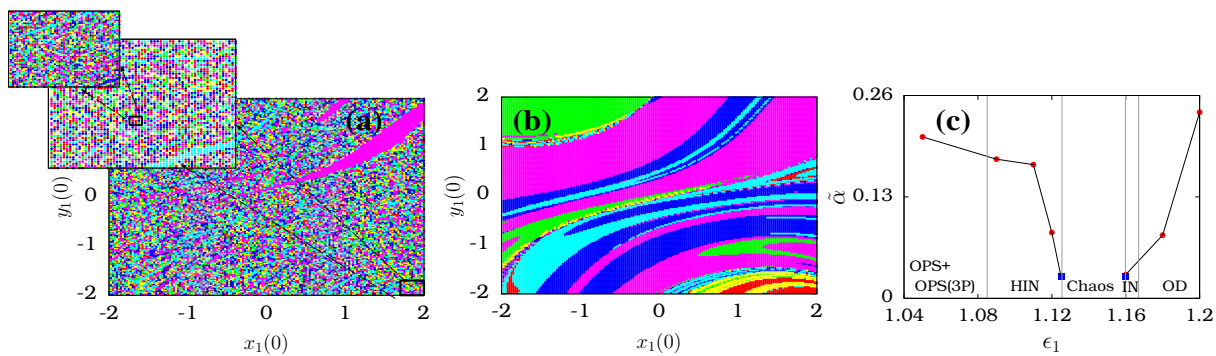}
	\caption{Figs (a) and (b) show the basin of attraction for $\epsilon_1=1.16$ (region of IN state) and $\epsilon_1=1.125$ (region OD state), respectively, for $\epsilon_2=1.7$.  The other state variables are initially fixed as $x_2(0)=1.1$, $y_2(0)=0.8$, $x_3(0)=0.7$, $y_3(0)=-0.3$.  In both the cases of IN and OD states, there are totally six possible configurations each of which is represented by different colors.  Fig. (c) shows the uncertainty exponent ($\tilde{\alpha}$) as a function of $\epsilon_1$  for fixed $\epsilon_2=1.7$. }
	\label{base}
\end{figure}
\par The observed transient chaotic regions ($I_1$, $I_3$, $S_5$ and OD regions) are multistable.  The erratic transient feature along with the multistable nature results in interesting intertwining between the basins of attraction of the different states as can be seen in the form of fractal basins or in the form of extreme type of fractal basin, namely, the riddled basin \cite{fractal, uncert, uncert2}.   For example, the basin of attraction for $\epsilon_1=1.16$ and $\epsilon_2=1.7$ is depicted in Fig. \ref{base}(a) and its subsequent enlarged figures show a closer view of the basin.  Figure \ref{base}(a) and its subsets corroborate the riddled form of the basin where the strong intertwining among the attractors is evident.  We have also displayed the basin of attraction of the OD states for $\epsilon_1=1.125$ and $\epsilon_2 = 1.7$  in Fig. \ref{base}(b) which shows the fractal nature of the basin.  The dynamics of the system in these fractal and riddled basins are also illustrated Appendix A.4..  Further in order to corroborate and quantify the fractal nature of the basin, the uncertainty exponent ($\tilde{\alpha}$) is calculated for different  vales of $\epsilon_1$ and is shown in Fig. \ref{base}(c) for a fixed $\epsilon_2$ (the procedure for calculating $\tilde{\alpha}$ is discussed  in Appendix A.5.). Note that  $\tilde{\alpha}$ is related to the probability that two nearby initial conditions lead to different attractors and $\tilde{\alpha} <1$ denotes the fractal nature of the basin while  $\tilde{\alpha} \approx 0$ denotes the riddled basins \cite{uncert, uncert2}.  The basins are riddled very near to the chaotic itinerancy region as confirmed by the values of the uncertainty exponent $\tilde{\alpha}$ in Fig. \ref{base}(c), revealing the high sensitivity of final asymptotic states with respect to the initial conditions.  Further, the sensitivity to the initial conditions reduces as $\epsilon_1$ deviates from the chaotic region ($I_2$) as is evident from the value of $\tilde{\alpha}$ with their corresponding basins as fractal basins (see Fig. \ref{base}(b)).  Thus it is strongly corroborated  that even coupled limit cycle oscillators can show transient chaotic behaviors that results in fractal or riddled basins.  The emergence of transient chaos around chaotic itinerant region has not yet been reported in the literature to the best of our knowledge.  Such a type of transient chaotic states around the chaotic itinerancy regions allow us to inherit both the  converging dynamics and chaotic wandering dynamics that are essential for achieving cognitive tasks and brain inspired computing paradigms as pointed out in the introduction.  This is because that in certain situations it is required that the system to remain in a particular memory state represented by a limit cycle or steady state (until the arrival of external stimuli) and in certain situations, the system is required to transit between different associated memory states to do sequential tasks or sequential memory recalling.  Hence, it is highly desirable and important to inherit both the type of dynamics in a simple system of coupled limit cycle oscillators. 
\subsection{Dynamics in  large networks}  
\begin{figure}[htb!]
	\centering
	\includegraphics[width=0.7\linewidth]{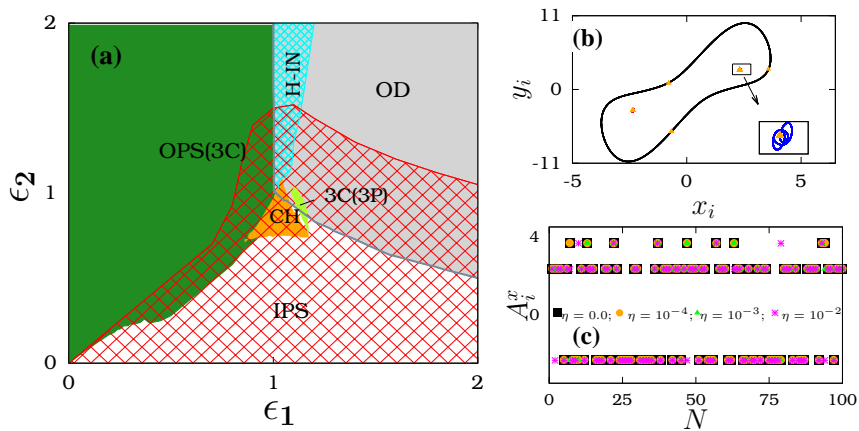}
	\caption{(a): The stable regions of different dynamical states of the globally coupled oscillators with $N=100$, where CH denotes the stable chaotic state. Fig. (b): The dynamics of the system in the H-IN region for $\epsilon_1=1.3$ and $\epsilon_2=3.0$ and Fig. (c) is plotted for the same parametric values where the amplitudes of oscillation $A^x_i$ is plotted with respect to the oscillator index for four different closely spaced initial conditions defined in the text. The stable boundaries of IPS and OD are the  obtained through  analytically from master stability function and	linear stability analysis.}
	\label{glob}
\end{figure}
\par Systems with increasingly large number of dynamical states are more desirable than that with a few number of states in view of the larger multistability required  to represent different operations or memory states.  Thus the search for the onset of transient chaos and fractal basins with more number of dynamical states is extended to a network of coupled vdP oscillators.  Stable regions of different dynamical states are depicted in Fig. \ref{glob}(a)  for globally coupled vdP oscillators with $N=100$.   The OPS, IPS and OD states exist even in the case of $N=100$ oscillators.  Segregation of oscillators into three clusters is observed in the OPS state (represented as OPS(3C)) and the three periodic OPS state in three clusters is represented as 3C(3P) in Fig. \ref{glob}(a).  Stable chaos is observed in the parametric range where super persistent transient chaos exists for $N=3$.  We found that an increase in the size of the network leads to its stabilization which is illustrated in Appendix A.6. It is also to be noted that the stabilization of chaotic attractor in this region does not affect the stability of synchronized periodic solution.  Thus interesting multistability between the completely ordered and completely disordered or chaotic states is observed around  equal attractive and repulsive couplings $\epsilon_1 \approx \epsilon_2 \approx 1$.   Existence of chaos due to the balanced excitatory and inhibitory couplings was one of the interesting research topics in the past \cite{natu}.    
\par Now considering the chaotic itinerant region and the transient chaotic dynamics, one can observe that the chaotic itinerancy is absent for $N>3$ but the transient chaotic nature prevails in all the cases which intertwines the homogeneous and inhomogeneous periodic orbits.  In the region denoted by H-IN in Fig. \ref{glob}(a), we observe multi-clusters with a few of the oscillators distributed in the homogeneous oscillatory branch (with the center at zero) and the others at the inhomogeneous oscillatory branches. For instance, Fig. \ref{glob}(b) shows a five cluster state for $\epsilon_1=1.3$ and $\epsilon_2=3.0$ where the three groups of oscillators evolve in the homogeneous periodic orbit with $\frac{2 \pi}{3}$ phase difference while  the other two groups are distributed in the inhomogeneous orbits.  Similarly, for certain parametric ranges, seven clusters can be observed where two more groups occupy the inhomogeneous orbits and so evolution of the  two groups of oscillators with $\pi$ phase difference can be observed in each of the  inhomogeneous orbits.   In these regions, the transient chaotic behavior intertwines the basin.  When we consider the H-IN region, the basins are found to be riddled and  show more sensitivity towards initial conditions.  However the sensitivity towards the initial condition is reduced when we consider the  H-IN state that emerge for higher values of $\epsilon_2$.  For example, in Fig. \ref{glob}(c), we have shown the distribution of oscillators among the homogeneous and inhomogeneous limit cycles for four different initial conditions.  The initial conditions are chosen as $x_i(0)=\zeta_i+\eta \tilde{\zeta}_i$, where $\zeta_i$'s are distributed randomly between $-1.0$ to $1.0$ and $\tilde{\zeta}_i$'s between $0$ and $1$.  $\eta$ decides the deviation of the initial conditions and we have chosen $\eta=0, 10^{-4}, 10^{-3}, 10^{-2}$.  The figure shows that when $\eta$ is of the order of $10^{-4}$ (or even lesser), there are no variations in the separation of oscillators into different orbits or no variation in the asymptotic state. But when the perturbation is  increased to the  order of  $\eta=10^{-3}$ and $10^{-2}$, there are deviations in the asymptotic state mimicking the fractal nature of the basins.  Such fractal nature may not  present in the H-IN state that exist for lower values of $\epsilon_2$ where the basins are found to be riddled.  But considering the OD states that are present near the boundary of H-IN state, their basins are found to be of fractal nature.  Thus it is interesting to note that we have not only shown the possibility to have fractal basin structures in oscillatory states but also in the OD state.

\subsubsection{Stability analysis}
The stable boundary for synchronized region obtained through master stability analysis \cite{msf1,msf3,msf2}.
In the synchronized manifold,  $x_i=x$,  and $y_i=y$, $\forall i$,   then the    variational equation of system (\ref{model}) can be written as, 
\begin{eqnarray}
{\dot{\eta}_{1j}}  &=&  \eta_{2j} +\epsilon_1\lambda_j\eta_{1j}, \qquad j=1,2,3,...N, \nonumber \\ 
{\dot{\eta}_{2j}} &=& -(2\alpha x_j y_j+1)\eta_{1j} +\alpha(1-x_1^2)\eta_{2j}-\epsilon \lambda_j\eta_{2j}.
\label{var}
\end{eqnarray}

where $x(t)$ and $y(t)$ are the solutions for the uncoupled  system  (\ref{model}).  ${\eta_{ij}}$ $(i=1,2)$ is the perturbation from the synchronized manifold, where  $\eta_{ij}={\boldsymbol {\beta_{ij} K_j}}$. Here $\boldsymbol{{\beta_{ij}}}=(\beta_{i1},\beta_{i2},...,\beta_{iN})$ and $\boldsymbol{{(\beta_{1j}, \beta_{2j})}}$ are the deviation of ($x_j$, $y_j$) from the  synchronized solution $(x, y)$.
$\boldsymbol{K_j}$ is the  eigenvector of the coupling matrix corresponding to the eigenvalue $\lambda_j$. The coupling matrix $\boldsymbol{K_j}$ is having the elements $a_{ij}={-(N-1)}/{N}$ for $i=j$ and $a_{ij}=a_{ji}={1}/{N}$ for $i\ne j$,  $(j=1,2,...N)$. To find  the stable region, we obtain the eigenvalues  from the coupling matrix ${\boldsymbol K_j}$, which is $\lambda_0=0$ and $\lambda_k=-1$, $k=1,2,...N-1$.  The  eigenvalue  $\lambda_0=0$ implies the perturbation parallel to the synchronization manifold and the remaining $N-1$ eigenvalues designates the perturbations transverse to the synchronization manifold.  In  the stable synchronization manifold the transverse eigenmodes should be damped out.  By substituting the eigenvalues  in  Eq.~\ref{var} and by finding the largest  Lyapunov  exponents the  stable boundary of the synchronized region is obtained and is  shown in Fig.~\ref{glob}.

Further, using linear stability analysis, we can find the stable boundary of the oscillation death state. In this case, the  oscillators in the network tend to an anti-symmetric inhomogeneous steady states which are of the form
\begin{eqnarray}
x^*=\pm \frac{\sqrt{1+2(p-1)\alpha\epsilon_1-4(p-1)^2\epsilon_1\epsilon_2}}{\sqrt{2}\sqrt{(p-1)\alpha\epsilon_1}}, \qquad   \quad  y^*=2\epsilon_1(p-1)x^*
\end{eqnarray}
The corresponding eigen values are, 
\begin{eqnarray}
\lambda_{1,2} = \frac{1}{2} \bigg(d_1\pm \sqrt{d_2}\bigg),  \quad   \lambda_{3,4} = \frac{1}{2} \bigg(d_1+2(p-1)(\epsilon_1-\epsilon_2)\pm\sqrt{d_3}\bigg)
\end{eqnarray}
where, $d_1=(1-x_1^2)\alpha$,  $d_2=\alpha^2-4(1+2x_1y_1)+x_1^2\alpha^2(x_1^2-2)$ and $d_3=d_1+2(p-1)(\epsilon_1-\epsilon_2)^2+4(2(p-1)(x_1^2-1)\alpha\epsilon_1-1-2x_1y_1\alpha+4(p-1)^2\epsilon_1\epsilon_2)$. The stable region of OD state emerges when $\epsilon_1\ge 1/(4(p-1)^2\epsilon_2)$ when $\epsilon_2< 1$ and $\epsilon_1\ge 1/(\sqrt{4(1-2p-p^2)})$ when  $\epsilon_2> 1$. The stable boundary for OD region is shown in Fig.~\ref{glob} for $p=0.5$.  The analytical  regions of synchronized state and oscillation death states   match well with the  numerical boundaries.  \\

\section{Conclusion}
\label{S:3}
The present report unraveled the possibility of onset of transient chaos and chaotic wandering even in a simple network of autonomous limit cycle oscillators.  We have elucidated that the transient chaotic nature that stems from frustration acts as a source for this type of dynamical behavior.   The transient chaotic nature in combination with the frustration induced multistability gives rise to fractal and riddled basin structures.  It is interesting to note that the complex basin structures have not only been observed in the case of oscillatory states but also in the case of oscillation death states.  Further, we have shown the existence of finite time transient chaotic dynamics in parametric regions around the chaotic itinerancy region including the existence of fractal and riddled basins for the first time in the literature using coupled limit cycle oscillators. The observed transient chaotic and chaotic itinerant region may find applications in the field of neuro-computation. However, the chaotic itinerancy behavior can be seen only in the three coupled system and it can not be observed in the case of larger networks with global coupling.  But the transient chaotic dynamics and riddled or fractal basin structures are observed even in  larger networks.   In addition, we have also shown the existence of super persistent transient chaos in the region $\epsilon_1 \approx \epsilon_2 \approx 1$ in smaller networks and it turns into stable chaos with an increase in the size of the network.  We firmly believe that large number of states with extreme multistability can be observed even in such a simple system of  limit cycle oscillators with suitable coupling which we plan to extend in future in view of its applications pointed out in the introduction. 
\section*{Acknowledgments}
\par KS sincerely thanks the CSIR for fellowship under SRF Scheme (09/1095(0037)/18-EMR-I).  SK thanks the Department of Science and Technology (DST), Government of India, for providing a INSPIRE Fellowship. The work of VKC forms part of a research project sponsored
by INSA Young Scientist Project under Grant No. SP/YSP/96/2014 and SERB-DST Fast Track scheme for
young scientists under Grant No. YSS/2014/000175. DVS is supported by the CSIR EMR Grant No. 03(1400)/17/EMR-II. The work of ML is supported by DST-SERB Distinguished Fellowship.  \\

\appendix
\section{}
\subsection{ \label{Appen:A} $0-1$ test to check transient chaotic nature}
\par Regular and chaotic nature of a deterministic system can be tested using the $0-1$ test \cite{gop_anna}.  We consider a time series of length $N$ corresponding to the transient region to test chaotic nature of the transient dynamics using $0-1$ test.  The first step along this direction is to compute the translation variables, $p(n)$ and $q(n)$, given as
\begin{figure}[h!]
	\centering
	\includegraphics[width=1.0\columnwidth]{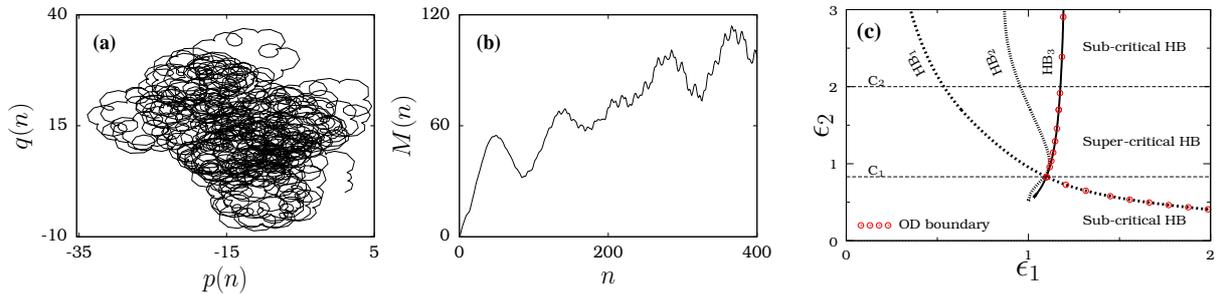}
	\caption{Fig. (a): Dynamics of translational variables $p(n)$ and $q(n)$ given in Eq. (\ref{cc}), Fig. (b): Dynamics of $M(n)$ with respect to $n$ and (c) Different Hopf bifurcation points are plotted in the $\epsilon_1$-$\epsilon_2$ space.  The points HB$_1$, HB$_2$ are indicated by pink diamonds in Fig.2(a) of the main text. The supercritical Hopf bifurcation point HB$_3$ is found to occur at the boundary between $S_2$ and $S_3$ (mentioned in Fig. 2(a) of the main text) where the stable inhomogeneous limit cycles bifurcating from this HB$_3$ point are represented by purple and green cross (X) marks.} 
	\label{zero}	
\end{figure}
\begin{eqnarray}
	p(n)=\sum_{j=1}^{n} x(j) \cos(jc), \qquad q(n)=\sum_{j=1}^{n} x(j) \sin(jc), \qquad n=1,2,3,...,N.
	\label{cc}
\end{eqnarray}

In the above, $x(j)$ represents the time series taken during the transient time.  The parameter $c$ in Eq. (\ref{cc}) can have any value between $0$ to $2 \pi$.  We have taken $c=0.7$ and $N=50,000$.  We have plotted the dynamics of the translational variables in Fig. \ref{zero}(a), where we have ensured that the transient dynamics prevails up to $N=50,000$.   Brownian like motion observed in Fig. \ref{zero}(a) corroborates that the transient dynamics is chaotic.  Further, the chaotic nature of the transient dynamics can also be confirmed by studying the diffusive or non-diffusive nature of the translational variables where the latter can be identified using mean square displacement $M(n)$,
\begin{eqnarray}
	M(n)=\underset{N \rightarrow \infty}{lim} \frac{1}{N} \sum \limits_{k=1}^N \left[p(k+n)-p(k)\right]^2+\left[q(k+n)-q(k)\right]^2.
	\label{mm}
\end{eqnarray}
The quantity $M(n)$ becomes bounded when the dynamics is regular and it becomes unbounded when the dynamics is irregular.  The behavior of $M(n)$ corresponding to the transient dynamics is given in Fig. \ref{zero}(b) and the unbounded nature observed in the figure clearly confirms that the transient dynamics is chaotic.  
\subsection{ \label{Appen:B} Hopf bifurcation curves}

Figure \ref{zero}(c) illustrates the Hopf bifurcation points in the $\epsilon_1$-$\epsilon_2$ space. The figure shows the three different Hopf bifurcation points, HB$_1$, HB$_2$, HB$_3$.  The inhomogeneous steady states are stabilized after one of these Hopf bifurcation points.  For values of $\epsilon_2$ lying the under line $C_1$, the stabilization of inhomogeneous steady states is found to occur via sub-critical Hopf bifurcation HB$_1$ (as observed in the case of two coupled systems).  Increasing $\epsilon_2$, the OD transition occurs via the super-critical Hopf bifurcation (HB$_3$). The latter bifurcation is responsible for the inhomogeneous oscillations that are observed in the IN region and the frustrated chaos in the region $I_2$.  As shown in Fig. \ref{zero}(c), above the line $C_2$, the stabilization of the OD state occurs via the sub-critical Hopf bifurcation so that the IN state and frustration induced chaos are suppressed with the increase of $\epsilon_2$ beyond $C_2$.

\subsection{ \label{Appen:C} Transient chaos induced by crisis: Mean exit time scale}
\par As mentioned in the main text,  destabilization of the chaotic orbits in the region $I_2$ does not seem to follow any particular bifurcation route other than self manifestation of it as transient chaos.  Such a transient chaotic state preceded by stable chaos is often referred(occur) as(via) boundary crisis in which the collision of chaotic attractor and unstable periodic attractor or its stable manifold occurs \cite{praa, scho2, pik}. To elucidate the above, we have calculated the mean exit time scales in the transient chaotic region.   Suppose that $\epsilon_1^*$ is the point at which the destabilization of the chaotic attractor occurs and the chaotic attractor is found to be stable in the region $\epsilon_1< \epsilon_1^*$ (or equivalently $\epsilon_1> \epsilon_1^*$).  In the region $\epsilon_1>\epsilon_1^*$ (or equivalently $\epsilon_1<\epsilon_1^*$), transient chaotic evolution prevails where the system behaves chaotically upto a time $\tau$ and then it reemerges from the  chaotic attractor to a stable periodic or point attractor.  The mean exit time scale ($\langle \tau \rangle$) can be obtained by averaging $\tau$ for different initial conditions.  In a variety of dynamical systems, it has been illustrated that this mean exit time scale follows a power law relationship for the values of $\epsilon_1$ near to the critical value $\epsilon_1^*$ as

\begin{eqnarray}
	\langle \tau \rangle = {\nu} (|\epsilon_1-\epsilon_1^*|)^{-\gamma}. \label{powlaw}
\end{eqnarray}

In the above, $\gamma$ is often known as critical exponent and it depends on the stability properties of the basic periodic orbit \cite{newpre, npre}.
\begin{figure}[h!]
	\centering
	\includegraphics[width=1.0\linewidth]{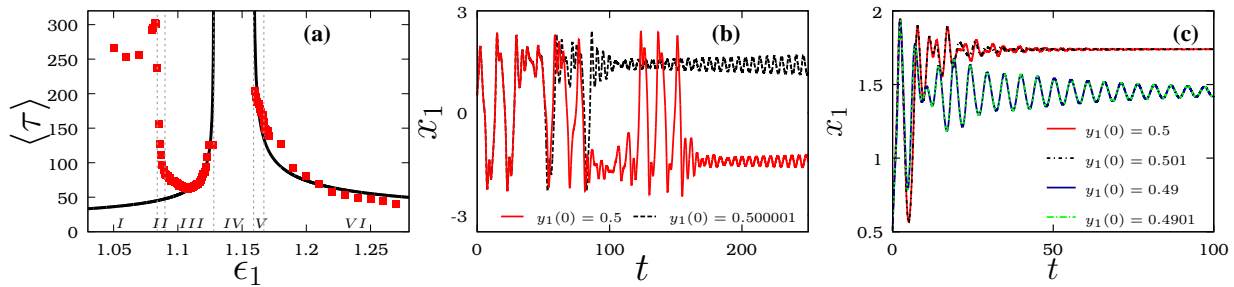}
	\caption{(a) Mean exit time scale calculated for different values of $\epsilon_1$ and for $\epsilon_2=1.7$.  The regions $I$ - $VI$ represent the stable regions of the following states: $I$ $\rightarrow$ OPS+OPS(3P) (region $S_5$ given in  Fig. 1(c) of the main article), $II$ $\rightarrow$ OPS+HIN, $III$ $\rightarrow$ HIN (regions $II$ and $III$ belong to $I_1$ region in  Fig. 1(c) of the main article), $IV$ $\rightarrow$ chaos (region $I_2$ in  Fig. 1(c) of the main article), $V$ $\rightarrow$ IN state (region $I_3$), $VI$ $\rightarrow$ OD state.  The mean exit time scales are calculated by considering $5000$ initial conditions and are represented by the filled squares. The obtained data is fitted to power law and the fit is denoted by black curve.  (b)  Dynamics in the riddled basin region ($\epsilon_1=1.16$, $\epsilon_2=1.7$) where the dynamics of the system corresponding to two closer initial conditions (a finite variation is given in $y_1(0)$) are shown.  (c) Dynamics in the fractal basin region where the dynamics of the system corresponding to two closer initial conditions (variation is given only in $y_1(0)$) are shown.  In this fractal region ($\epsilon_1=1.19$, $\epsilon_2=1.7$ which corresponds to OD region), even though the dynamics show sensitivity towards initial conditions, it is not too high.   } 
	\label{bnn}	
\end{figure}
\par To understand the destabilization of chaos due to boundary crisis, the mean exit time is plotted for different values of $\epsilon_1$ in Fig. \ref{bnn}(a) for $\epsilon_2=1.7$.  The figure shows that the chaotic dynamics prevails in the region $IV$, $1.128 \leq \epsilon_1 \leq 1.159$, and so the mean exit time blows up to infinity at the two critical points $\epsilon_1^*=1.128$ and $\epsilon_1^*=1.159$.  Before ($\epsilon_1<1.128$) and after ($\epsilon_1>1.159$) this chaotic region, the transient chaotic behavior is found to exist.   First, let us consider the chaotic destabilization at the boundary between regions $IV$ and $V$ or at $\epsilon_1^*=1.159$.  At this critical point $\epsilon_1^*=1.159$, the collision of chaotic attractor with the inhomogeneous periodic orbits corresponding to IN state gives rise to chaotic saddles and so we observe the finite time chaotic evolutions. When the IN state becomes more stable, the widening of the IN state basin occurs and it results in the decrease of mean exit time for the values of $\epsilon_1>1.159$.  By fitting the data to power law form as given in Eq. (\ref{powlaw}), we have found the critical exponent as $\gamma=0.367$. 
\par Similarly, when considering the transient chaotic regime present below $\epsilon_1^*=1.128$, we found that the mean exit time given in Fig. \ref{bnn}(a) shows power law  for the values of $\epsilon_1$ near to $\epsilon_1^*=1.128$ and not for the ones lying away from $\epsilon_1^*$.   When $\epsilon_1$ deviates from $\epsilon_1^*=1.128$, the mean exit time shows sudden increase in its value and attains a maximum near the boundary between regions $I$ and $II$.  However, in the region $I$, the mean exit time again  decreases. Such an absence of monotonic decrease in the mean exit time scale has also been reported in \cite{newpre}.  Here, at the critical value $\epsilon_1^*=1.128$, the collision between the chaotic attractor with the periodic orbit corresponding to HIN state results in the transient chaotic behavior.  As $\epsilon_1$ deviates from $\epsilon_1^*$, the HIN state  become more stabilized resulting in widening of their basin  so that the decrease in the mean exit time occurs for the values of $\epsilon_1$ near to $\epsilon_1^*$.  But for the values of $\epsilon_1$ lying near to the boundary between region $III$ and $II$ and also in the region $II$, the HIN state may become weak and paves way for the stable OPS and OPS(3P) states.  Due to the above fact, we may observe an increase in the value of mean exit time.  However after the stabilization of OPS(3P) state in the region $I$ and due to the increase in the stable nature of the OPS and OPS(3P) states, the mean exit time starts decreasing in region $I$.  Due to this fact, the mean exit time follows power law  relationship only near $\epsilon_1^*=1.128$ as  shown in Fig. \ref{bnn}(a) corroborating  the destabilization of chaotic orbit due to the boundary crisis at both the ends of the chaotic region (region - $IV$). 

\subsection{ \label{Appen:D} Dynamics in riddled and fractal basin regions}

\par The sensitivity of the system towards initial conditions in the riddled and fractal basin regions is illustrated in Figs. \ref{bnn}(b) and \ref{bnn}(c) which are plotted for $\epsilon_1=1.16$ and $\epsilon_1=1.19$, respectively.  By considering two different initial conditions that differ in $y_1(0)$ by $10^{-6}$ (and keeping other initial condition values fixed), we have plotted their dynamics in Fig. \ref{bnn}(b).  The latter shows that the trajectories corresponding to two initial conditions are close only up to a finite time.  After that, the two trajectories rapidly deviate from each other and evolve towards different attractors. But in the fractal basin, the system does not show extreme sensitivity which is obvious from Fig. \ref{bnn}(c).  The closer initial conditions lead to  the same  attractor state whereas well separated initial conditions lead to different attractors.

\subsection{ \label{Appen:E} Uncertainty exponent}
\par  Both fractal and riddled basins correspond to non-smooth basin boundaries that make sensitive dependence of final states towards initial conditions.   Compared to fractal basins, the riddled ones show extreme sensitivity towards initial conditions.  Considering the case of fractal basins, the probability to predict the asymptotic state is improved when the initial condition is more precisely defined.  In the case of riddled basins, the final state is found to be uncertain even if the initial condition is precisely defined.  These two types of basins can be classified  by calculating the uncertainty exponent \cite{awadhesh2}.  The latter is related to the probability that the two nearby initial conditions lead to same asymptotic state or not.  For this purpose, the attractor reached by a set of randomly distributed or equally spaced initial conditions $\vec{X}_i(0)(=(x_i(0), y_i(0))$) has to be determined and then one has to determine the asymptotic state of another set of initial condition which is deviated from the earlier set by a finite value $\epsilon_i$ as $\vec{X}_i(0)+\vec{\epsilon}_i$.   The probability ($P(\epsilon)$) that this pair of nearby initial conditions ($\vec{X}_i(0)$ and $\vec{X}_i(0)+\vec{\epsilon}_i$) leading to different asymptotic states is computed for different values of separation (${\epsilon}$).  The uncertainty exponent is determined by fitting the scaling law $P(\epsilon) \sim \epsilon^{\tilde{\alpha}}$ where $\tilde{\alpha}$ represents the uncertainty exponent.  For riddled basins $\tilde{\alpha} \approx 0$.
\begin{figure}[h!]
	\centering
	\includegraphics[width=0.8\linewidth]{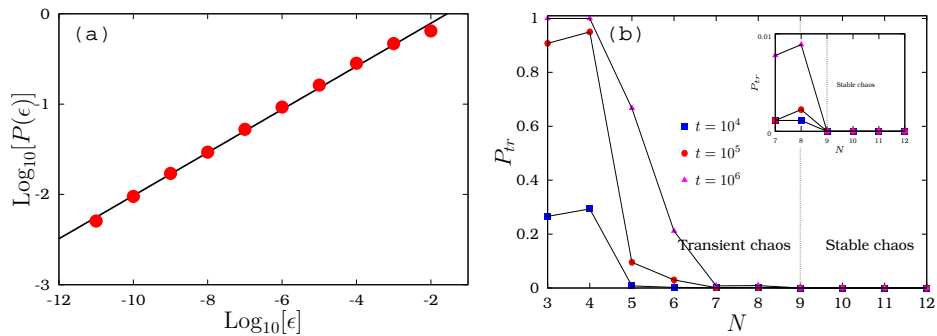}
	\caption{ (a) Figure is plotted in the OD region where $\mathrm{Log}_{10}[P(\epsilon)]$ is plotted for $\epsilon$.  It is fitted to a straight line and the slope of this straight line gives the value of uncertainty exponent, which  turns out to be $\tilde{\alpha}=0.238561$. (b) Transition from transient chaos to stable chaos with the increase in the number of oscillators.  Here we have calculated probability  $P_{tr}$ at different time instants $t=10^4, 10^5, 10^6$ that whether the system reaches the asymptotic state or not.  When $N<9$, $P_{tr}$ increases with time and so the chaotic behavior is a transient one.  When $N \geq 9$, we observe that for whatever value of $t$, the probability remains zero elucidating the stabilization of chaos. } 
	\label{l10}	
\end{figure}
\par  The probability of the system to reach the same attractor for a set of initial conditions that differ by length $\epsilon$ is shown Fig. \ref{l10}(a).  The slope of the straight line shown in Fig. \ref{l10}(a) gives the value of the uncertainty exponent $\tilde{\alpha}$.

\subsection{\label{Appen:F} Transient chaos to stable chaos}

\par As seen in the main article, the system exhibits long persisting transient chaos for $N=3$ in the region where the value of the attractive and repulsive couplings are approximately unity.  The transient chaos turns out to be stable chaos when the size of the network is increased.  To illustrate the above, we have plotted the probability of the system to reach the periodic state ($P_{tr}$) at a particular time in Fig. \ref{l10}(b).  For different values of $N$, $P_{tr}$ is calculated at times $t=10^4,10^5$ and $10^6$ by considering different initial conditions.  For $N=3$ and $t=10^4$, we observe that the system remains in the chaotic state for many of the initial conditions and so $P_{tr}$ takes finite values whereas for $t=10^6$, the system reaches the periodic state for almost all the initial conditions so that $P_{tr} \approx 1$.  By increasing $N$ (upto $N=9$), the lifetime of transient chaos is also increased so that $P_{tr}$ takes up finite values even at $t=10^6$.   $P_{tr}$ becomes zero for $N>9$ in Fig. \ref{l10}(b) which indicates the stabilization of chaotic state.

%

%
\end{document}